\documentclass[preprint2]{aastex}
\usepackage{graphicx}
\usepackage{natbib}

\shorttitle{Short title}
\shortauthors{Author}

\begin{document}

   \title {Interpretation of Solar Irradiance Monitor measurements
through analysis of 3D MHD simulations.}
   \author{S.~Criscuoli\altaffilmark{1}, H.~Uitenbroek\altaffilmark{1}}

\altaffiltext{1}{ National Solar Observatory, Sacramento Peak, 
P.O.\ Box 62, Sunpsot, NM 88349, USA}

   \date{}

\begin{abstract}
Measurements from the Solar Irradiance Monitor (SIM) onboard the SORCE
mission indicate that solar spectral irradiance at Visible and IR
wavelengths varies in counter phase with the solar activity cycle. The
sign of these variations is not reproduced by most of the irradiance
reconstruction
techniques based on variations of surface magnetism employed so far, and it is not clear yet whether SIM
calibration procedures need to be improved, or if instead new physical
mechanisms must be invoked to explain such variations. We employ
three-dimensional magneto hydrodynamic simulations of the solar
photosphere to investigate the dependence of solar radiance in SIM
Visible and IR spectral ranges on variations of the filling
factor of surface magnetic fields.
We find that the contribution of magnetic features to solar radiance is strongly dependent on the location on the disk of the features, being negative close to disk center and positive toward the limb. If features are homogeneously distributed over a region around the equator (activity belt) then their contribution to irradiance is positive with respect to the contribution of HD snapshots, but decreases with the increase of their magnetic flux for average magnetic flux larger than 50 G in at least two of the Visible and IR spectral bands monitored by SIM. Under the assumption that the 50 G snapshots are representative of quiet Sun regions we find thus that the Spectral Irradiance can be in counter-phase with the solar magnetic activity cycle. 
\end{abstract}


\maketitle

\section{Introduction}
Solar irradiance, the radiative energy flux the Earth receives from the Sun 
at its average orbital distance, varies along with magnetic activity,
over periods of days to centuries and presumably even on longer time scales.
The magnitude of irradiance variations strongly depends on wavelength.
The precise measurement of irradiance over the spectrum
is becoming more and more compelling, because of the increasing evidence of the effects of these variations on the chemistry of the \textbf{Earth's}
atmosphere and terrestrial climate \citep[e.g.][ and references
therein]{lockwood2012, ermolli2013}.
However, absolute measurement of spectral irradiance variations,
especially over time scales longer than a few solar rotations,
is seriously hampered by difficulties in 
determining degradation of instrumentation in space.
 Therefore, calibrations of radiometric measurements have to rely 
significantly on inter-calibration with other instruments and/or
reconstructions through models based on proxies of magnetic activity.

In this context, recent measurements obtained with the Spectral
Irradiance Monitor \citep[SIM;][]{harder2005} radiometers on board the
Solar Radiation and Climate Experiment \citep[SORCE;][]{rottman2005},
which show an irradiance signal at Visible and IR spectral bands in
\emph{counterphase} with the solar cycle \citep{harder2009}, have been
strongly debated. This result was confirmed by
\citet{preminger2011}, who found that variations in irradiance of solar and
solar-like stars in red and blue continuum band-passes
is in counter phase with their activity cycle.
By contrast, recent results obtained from the analysis of
\textbf{VIRGO/SOHO} \citep{frohlich1995} data at visible spectral ranges
\citep{wehrli2013} show signals \emph{in phase} with the magnetic
cycle.\textbf{ Theoretically, most of the irradiance reconstruction techniques
which usually
reproduce more than $90\%$ of variations of total solar irradiance
(i.e., the irradiance
integrated over the whole spectrum), produce irradiance variations
at SIM Visible and Infrared bands that are in phase with the magnetic
cycle \citep[see][for a review]{ermolli2013}. The Spectral and Total Irradiance REconstruction models for Satellite Era (SATIRE) produce a signal slightly in counter-phase in the IR \citep[e.g.][]{ball2011}. The only reconstructions that produce a signal in counter-phase with the magnetic activity cycle on both visible and IR bands are those obtained with the Solar Radiation Physical Modelling (SRPM) tools
\citep{fontenla2012}, which, on the other hand, have been criticized for being explicitly constructed to
reproduce SIM measurements.}

Given the above controversy it is still an open question whether SIM
finding of counter phase spectral variation in visible
and IR bands is the result of a problem with internal calibration
procedures
, or if instead
current modeling is not adequate in reproducing irradiance variations at
those spectral ranges. In particular the physical cause of
long-term variations is still unclear having been
attributed alternatively to changes in quiet Sun magnetism that is
mostly hidden in full-disk observations \citep{fontenla2012},
or to a change of the temperature gradient in the solar atmosphere,
most likely due to an increase of the magnetic filling factor over the cycle
\citep{harder2009}.

\textbf{Several irradiance reconstruction techniques, such as the SATIRE and the SRPM cited above, the reconstructions of the Astronomical Observatory of Rome \citep[OAR,][]{ermolli2011}, and those obtained with the Code for Solar Irradiance \citep[COSI,][]{haberreiter2008, shapiro2010} and with the Solar Modelling in 3D \citep[SOLMOD,][]{haberreiter2011}, are based on one-dimensional
static atmosphere models.} Such models can be constructed to
reproduce observed spectra very well, but their semi-empirical
nature prevents them from being used to explore the underlying physics \citep{uitenbroek2011}.
In this contribution we employ snapshots from 
3-D magneto-hydrodynamic (MHD) simulations of the solar
photosphere to qualitatively investigate whether an increase of the
magnetic filling factor over the solar surface can produce a decrease
of the disk-integrated solar radiative emission in the four visible
and IR spectral bands monitored by SIM. Since the contribution of
features as pores and sunspots is well known to be negative,
this study is aimed at investigating the contribution of features
like faculae and network.

The paper is organized as follows: in Sect. 2 we describe the MHD
snapshots employed and the spectral synthesis performed; in Sect. 3 we
present our results and in Sect. 4 we draw our conclusions.
\begin{figure*}[!] 
\includegraphics[width=16.5cm] {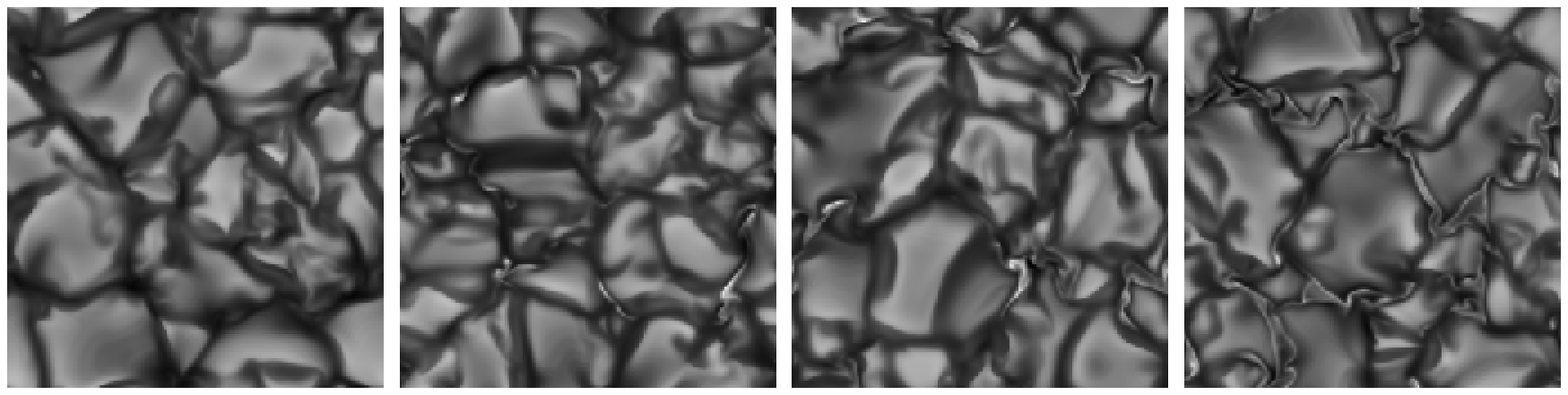}
\caption{Vertical line of sight emergent radiation at 630 nm through MHD snapshots characterized by different amounts of average magnetic flux. From left to right: HD, 50 G, 100 G and 200 G.}
\label{images} 
\end{figure*} 
\section{Simulations and Spectral Synthesis}
\label{data:simulations}
For our analysis we 
consider series of snapshots of 3-D MHD simulations of the solar
photosphere calculated  by \citet{fabbian2010, fabbian2012} using the STAGGER code \citep{garlsgaard1996}.
 They are
characterized by four different cases of introduced magnetic flux with
average vertical field strength values of approximately 0 G, 50 G, 100 G and 200 G.  The 0 G
case (hydrodynamic, or HD case hereafter) and the 50 G case are
representative of quiet Sun regions, while the 100 and 200 G are
meant to represent magnetic regions. The horizontal dimensions of each
snapshot are 6 Mm $\times$ 6 Mm, with a spatial sampling of 23.8 km
resulting in a grid size of 252 points in both horizontal directions.
These snapshots were used by Fabbian (2010, 2012) for the first-ever
quantitative assessment of the impact of magnetic fields in 3D
photospheric models of the Sun on the solar chemical composition. The
numerical setup adopted for the calculations is described in the above
cited papers.  
\citet{beck2013} discussed the quality of the simulations and
their comparison with observations. 
\textbf{ \citet{criscuoli2013} employed this set of simulations to investigate physical  and observational differences between quiet and facular regions. Among other results, this study showed that, in agreement with high-spatial-resolution observations,  the emergent intensity at disk center in the red continuum of magnetic features characterized by the same size and the same amount of average magnetic flux decreases with the increase of their environmental magnetic flux. That is, small-size magnetic features are brighter in quiet areas with respect to active regions. Here we extend that work to a larger set of wavelengths and lines of sight and investigate the effects of the magnetic flux on the radiative emission  as would be observed by moderate spatial resolution data (as those usually employed for irradiance studies). }
\textbf{ At this aim, we considered each of these snapshots to represent a patch of unresolved
magnetic field with the corresponding average flux}. At each pixel we calculated with
the RH code \citep{uitenbroek2003} the emergent radiation by solving
for LTE radiative transfer at 16 continuum wavelengths,
distributed equidistantly over 6 spectral positions in each of the SIM
pass bands (400 -- 972 nm, 972 -- 1630 nm, and 1630 -- 2423 nm). \textbf{Figure  \ref{images} illustrates examples of the emergent intensity along the vertical line of sight at  630 nm through snapshots characterized by different amounts of average magnetic flux}.  We
solved radiative transfer through the snaphots in different directions
spanning 9 inclinations, distributed according to the zeroes of the
Gauss--Legendre polynomials in $\mu = \cos(\theta)$, where $\theta$ is
the angle with the vertical, and 2 azimuths per octant, plus the
vertical direction, for a total of 73 directions.  We then averaged
over the different azimuths and all the spatial positions in the
snapshot to calculate the snapshot's average emergent intensity at all
16 wavelengths as a function of heliocentric angle.

\begin{figure}[!] 
\includegraphics[width=4.6cm, trim=1cm 1cm 7cm 0cm] {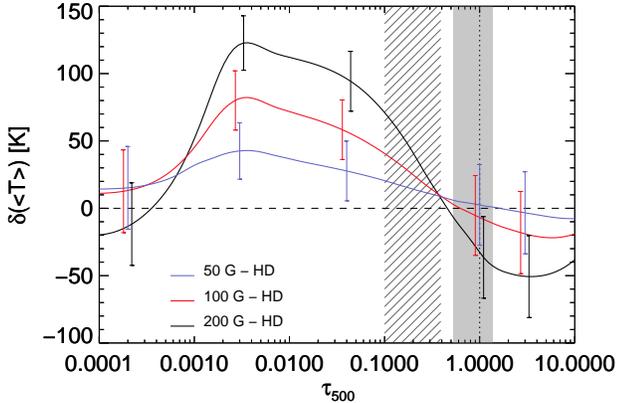}

\caption{Difference between MHD average temperature stratifications
  and the HD case versus the optical depth $\tau_{500}$.  Error bars represent the standard deviation obtained
  by averaging over the 10 snapshots each. The solid and dashed gray areas indicate the approximate formation heights for the wavelenghts considered for  vertical and $\mu$ =0.2 lines of sight, respectively.}

\label{gradient} 
\end{figure}

To ensure sufficient statistics and reduce oscillation effects,
for each average magnetic flux value we considered 10 snapshots 
taken 2.5 minutes apart (40 snapshots in total). 
Figure \ref{gradient} shows the differences
between the average temperature stratifications of the MHD and the HD
atmospheres as function of the optical depth computed at 500 nm.
These curves agree with those reported in \citet[][their figure
  1]{fabbian2010}, within the statistical variations due to here
considering fewer snapshots from the complete series, and to possible
slight differences in the computed optical depth due to employing a
different radiative transfer code. \textbf{The solid gray area approximately
indicates the range in formation heights of the continuum intensity
along a vertical line of sight at the considered wavelengths. The dashed gray area shows approximately the formation heigths at the same wavelengths for an inclined line of sight (namely, for $\mu$ = 0.2, where $\mu$ is the cosine of teh heliocentric angle); as expected, the shallower the line of sight, the more the formation heights shift toward higher layers of the atmosphere (smaller optical depth values).} The
plot shows that an increase of the average magnetic flux causes an
average decrease of the temperature at about optical depth unity and
an increase at smaller optical depths, thus making the temperature
gradient shallower with the increase of the magnetic flux. \textbf{As explained in detail in \citet{criscuoli2013}, these variations must be ascribed to the inhibition of convection by the magnetic field, which reduces the amount of energy transported by the plasma from the lower to the higher layers of the atmosphere.} These
temperature gradient changes are similar to those invoked by \citet{harder2009}
to explain SIM measurements at Visible and IR bands, and are
qualitatively similar to the differences between temperatures of quiet
and magnetic feature atmosphere models such as employed in SRPM
\citep[e.g.][]{fontenla2012}.  We note however, that because of our
inclusion of the true three-dimensional structure of the snapshots,
the centre-to-limb behavior of our computed intensities is more
realistic than that of the one-dimensional models, even though the
snapshots possess, on average, similar temperature gradients.

After calculating the spatially and azimuth averaged 
emergent intensities as a function of
$\mu$ for each snapshot we averaged them over the 10 realizations
for each average magnetic flux case, and compared the intensities
for each of the MHD cases with those of the HD case as reference.
Intensity contrast is, therefore, defined as the intensity relative to that
of the HD case at the same heliocentric angle: $C = <I_{MHD}>/<I_{HD}> - 1$,
where the averages are over spatial dimensions in each snapshot,
realization, and azimuth. Finally, we computed the flux
by integrating intensities over a range of heliocentric angles.

\begin{figure} 
\includegraphics[width=5.5cm,height=8.5cm, trim=1cm 1cm 7.5cm 1cm] {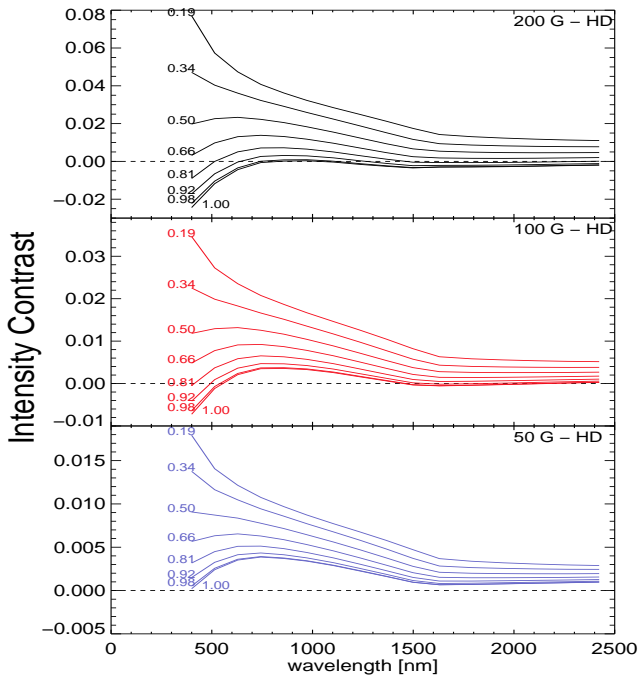}

\caption{Wavelength dependence of the average intensity contrast of
MHD models compared to the HD case for different heliocentric angles
(labeled by their cosine).}  

\label{CLVs} 
\end{figure} 

\section{Results}
The plot in Fig.\ \ref{CLVs} shows the average intensity contrast of the
MHD snapshots as a function of wavelength for different
heliocentric angles. 
We note that the intensity contrast is negative for the 100 G and 200 G
cases near disk center ($\mu > 0.8$) both for short wavelengths (below approximately 500 nm for the 100 G snapshots and below approximately 700 nm for the 200 G ones)
and wavelengths above 1500 nm, but that the contrast is positive
for all angles and wavelengths in the 50 G average field case.
This is because at short and long wavelengths intensity emanates
from relatively deep layers of the solar atmosphere, at which, as illustrated in Fig. \ref{gradient} \textbf{(continuous gray area)}, the 100 and 200 G simulations have average temperature lower than the one of the HD case.
\textbf{For shallower lines of sight ($\mu \leq$ 0.8) the formation heights shift
toward higher layers of the atmosphere (dashed gray area in Fig. \ref{gradient}), where the
difference between the average temperatures of MHD and HD
simulations becomes positive.}
These results are in agreement with
those obtained from observations
\citep[e.g.][]{yeo2013,ermolli2007,ortiz2002, sanchezcuberes2002}.
Close to disk center
most of the curves exhibit a maximum at about 800 nm.  This wavelength
dependence of the contrast closely follows the behavior of the $H^{-}$
opacity, which is the prominent source of opacity in the
photosphere, and has a maximum near that wavelength.
At those wavelengths intensity forms relatively high in the atmosphere,
where the MHD snapshots are on average warmer than the HD ones.
\begin{figure}
\includegraphics[width=5.5cm, height=8.2cm, trim=1cm 1cm 7.5cm 1cm]%
{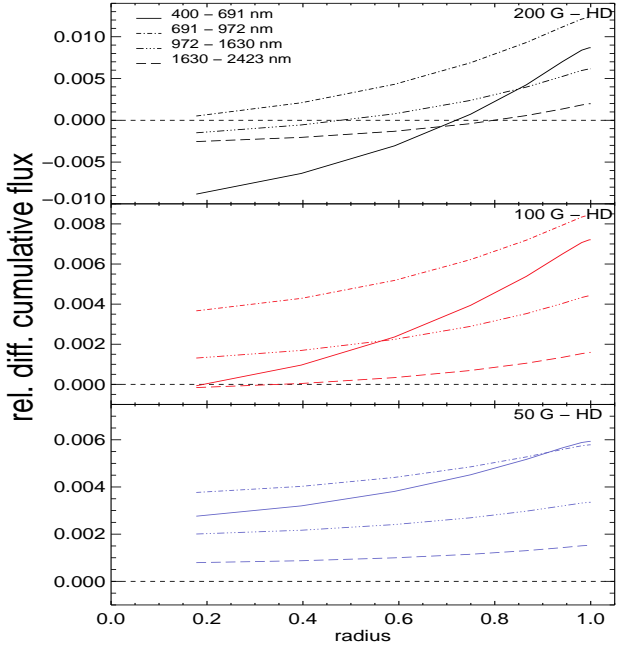}

\caption{Cumulative flux relative difference between MHD and HD models,
         integrated over SIM spectral bands. Cumulative flux
         is calculated by integrating intensity from disk center out
         to the specified radius on the x-axis.}
\label{fluxes} 
\end{figure} 
\begin{figure}
\includegraphics[width=5.5cm, trim=1cm 1cm 7.5cm 1cm] {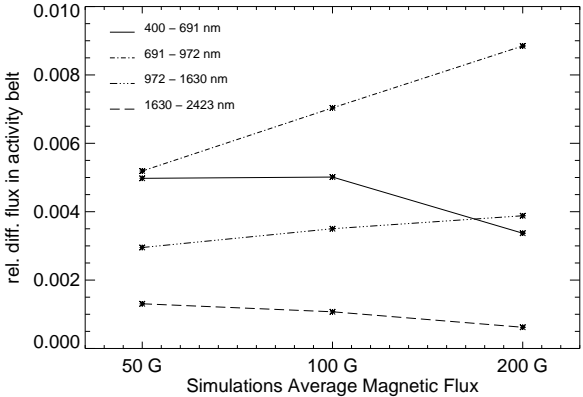}

\caption{Cumulative flux relative difference between MHD and HD snapshots computed over the activity belt \textbf{(defined here as the solar surface extending from + to - 30 degrees around the equator) } and integrated over SIM spectral bands.}

\label{fluxesactivitybelt} 
\end{figure} 

The derived intensity contrasts would suggest that it is in principle
possible to have a decrease of the irradiance at continuum
wavelengths due to the presence of magnetic features (other than
sunspots and pores), in particular when these have an average vertical
flux of at least 100 G, appear preferentially
near disk center, and are observed at short (below 600-700 nm), 
or long (above 1500 nm) wavelengths.
However, we have to verify that negative contrasts still could
appear when activity is more evenly spread out over the disk,
as is the case when averaging over several solar rotations,
and when integrated over the SIM wavelength bands.
We therefore estimated the cumulative radiative flux
computed over the surfaces of disks of increasing partial radius and
integrated these fluxes in wavelength over the four bands 
observed by SIM.
Figure \ref{fluxes} shows the relative differences between the cumulative
flux computed for the MHD
snapshots and the HD ones as function of partial solar radius.
The curves confirm that the contribution of faculae to the radiative
flux in SIM bands is always positive in the 50 G case, while it can
turn negative as the magnetic flux increases (in particular, this can
be seen for the curve corresponding to the 200 G simulation), if these
features are preferentially located close to disk
center. Nevertheless, the contribution is positive if these magnetic
features are uniformly distributed over the disk (cumulative radius of
one).  The main reason for this positive contrast is the limb
brightening of the intensity contrast that occurs for all wavelengths
(see Fig.\ref{CLVs}), and is the result of the sampling of higher
temperatures in the MHD snapshots at shallower viewing angles, and the
additional brightening that stems from geometrical effects when
shallow viewing directions sample hotter material behind partially
evacuated magnetic field concentrations. This latter effect cannot
adequately be represented by one-dimensional modeling.

Typically, magnetic activity occurs preferentially in activity belts
North and South of the equator, moving from higher latitudes in the
early phase of the solar cycle to lower ones in later phases.
Limiting magnetic elements to lower latitudes limits the number
of magnetic elements at higher heliocentric angles, potentially
allowing for a negative contribution to the irradiance.
To test this possibility we computed the disk integrated radiative flux of 
magnetic elements confined to an activity belt between latitudes
of $\pm$ 30 degrees.
Fig. \ref{fluxesactivitybelt} shows the result of this calculation
for the SIM bands. In particular, it shows the radiative flux relative differences between the MHD and HD snapshots integrated over the SIM bands
as function of the average vertical magnetic-flux. It clearly
shows a positive contrast for all field cases considered in all
four bands. Nevertheless, it also shows that in the 400-691 nm and in the 1630-2423 nm bands the radiative fluxes relative differences decrease with the increase of the magnetic flux. This suggests that, if magnetic features are preferentially located over the activity belt, and if the relative number of features with higher magnetic flux increases with the increase of the magnetic activity, then the radiance at those two SIM bands decreases.\textbf{ Moreover if, more "realistically", we consider that even the quiet Sun is permeated by magnetic flux \citep[see][for a review on quiet Sun magnetic field]{pillet2013} and we take as reference the 50 G snapshots, then the contribution of facular regions at those two SIM pass bands is always negative.  }


\section{Discussion and Conclusions}
We employed snapshots from 3-D MHD simulations, characterized by
different values of average vertical magnetic flux, to estimate solar
irradiance variations at the visible and IR spectral ranges of SIM
radiometers, stemming from contributions of patches of unresolved
magnetic field.  The results from our spectral synthesis confirm the
fact the contribution of facular region to irradiance is strongly
dependent on their location over the solar disk \citep[see also the
  discussion in][]{fontenla2012}.  In particular, we find that the
increase of the magnetic filling factor over the solar surface can
produce a {\it decrease} of emitted radiation only for mostly vertical
lines of sight and only for wavelengths below 500-700 nm (depending on
the magnetic flux), or above 1500 nm (Fig. \ref{CLVs}).  Integrating
the intensity over the disk, even if we limit the contribution of
magnetic regions to an activity belt, always renders the contribution
of the magnetic elements to the irradiance positive
(Fig. \ref{fluxesactivitybelt}). Nevertheless, if magnetic features
are distributed over the activity belt, their contribution decreases
at two of the SIM bands (namely at 400-691 and 1630-2423 nm) with the
increase of the average magnetic flux. This suggests that, assuming
that the relative number of features with larger magnetic flux
increases with the increase of the magnetic activity, then the
spectral irradiance at those SIM bands can decrease toward solar
maximum. 
\textbf{Results shown in Fig. \ref{fluxesactivitybelt} also indicate that, if we take as reference the 50 G snapshots instead of the HD ones, then the contribution of facular regions to irradiance at the 400-691 and 1630-2423 nm SIM wavelength bands is always negative. We note that this is a more "realistic" assumption than taking the HD snapshots as reference, as previous works have shown that  MHD simulations with average vertical magnetic field between 20 - 30 G best represent properties of magnetic field of the quiet Sun \citep{khomenko2005, danilovic2010}.  Finally,
we note that magnetic features tend to appear toward
higher latitudes at the beginning of the cycle, migrate toward the
equator  as
the magnetic activity peaks, and that then part of their flux,
fragmented into lower magnetic flux features, tends to migrate toward
the poles during the descending phase. Since, as we have shown, the contribution to irradiance of these features strongly depends on their position on the solar disk,  we speculate that multiple
peaks of the solar spectral irradiance could be observed.  }
   
Note that an average flux of 200 G, the maximum we considered,
is modest for facular regions, as values up to 800 G are usually
employed for reconstructions \citep[e.g.][and references therein]{ball2012}.
On the other hand, from results shown in this work as well as from results
obtained from numerical simulations by other authors
\citep[e.g.][]{vogler2005}, and from observations
\citep[e.g.][]{yeo2013, ortiz2002} it is clear that the center-to-limb
variation of contrast increases with magnetic flux so that it is likely
that our conclusions would be even stronger.

Likewise, the inclusion of spectral lines in our calculations
would have most likely increased the contrast between MHD and
HD intensities, as spectral lines contribute opacity and raise
the formation height of the spectral bands, causing them to
sample slightly higher layers, where the differences between
the average temperatures of the models with different field strength
is larger. Nevertheless, we expect this effect to be larger for the lower magnetic flux simulations, where the average temperature gradient is steeper (and spectral lines are deeper), with respect to higher magnetic flux simulations, thus increasing the steepness of the relations in Fig. \ref{fluxesactivitybelt} for the 400-691 ans 1630-2423 nm bands, and decreasing the steepness of the curves of the other two bands.  This effect too would thus strengthen our conclusions.

\textbf{We therefore conclude that the spectral synthesis presented in this study are  compatible with a negative contribution of facular regions to the irradiance
in the SIM visible and IR bands with an increase in magnetic filling
factor if as reference for quiet Sun we assume snapshots of 50 G average
magnetic flux. }

\acknowledgements
The snapshots of magneto-convection simulations were provided to us by
Elena Khomenko and were calculated using the computing resources of
the MareNostrum (BSC/CNS, Spain) and DEISA/HLRS (Germany)
supercomputer installations.

\end{document}